\documentclass[twocolumn,twoside,slac_two]{revtex4}
\usepackage{graphicx}
\usepackage{fancyhdr}
\begin{document}

\title{High resolution radio observations of gamma-ray emitting 
Narrow-Line Seyfert 1s}

%

\author{M. Orienti}
\affiliation{Astronomy Department, Bologna University, Italy; INAF-IRA Bologna, Italy}
\author{F. D'Ammando}
\affiliation{Physics Department, Perugia University, Italy; INAF-IRA Bologna, Italy}
\author{M. Giroletti}
\affiliation{INAF-IRA Bologna, Italy}
\author{on behalf of the Fermi Large Area Telescope Collaboration}

\begin{abstract}
The detection by {\it Fermi}-LAT of $\gamma$-ray emission from radio-loud
Narrow-Line Seyfert 1s (NLS1s) indicates that relativistic jets do not form
only in blazars and radio galaxies, but also in other AGN populations. 
Despite a spectral energy distribution
similar to blazars, their physical characteristics are quite different: lower
black hole masses, generally higher accretion rates, and possibly hosted in
spirals. Furthermore, their radio properties make the interpretation of these
objects even more puzzling. The radio emission is very compact, not exceeding
the parsec scales, as also found in the
population of young radio sources. We present high resolution VLBA
observations of three radio-loud NLS1s detected by {\it Fermi}-LAT: 
SBS\,0846+513, PKS\,1502+036, and PKS\,2004$-$447. The information on the 
pc-scale morphology will be complemented with studies of flux density 
and spectral variability from multi-epoch and multifrequency
observations, in order to unveil the nature of their radio emission. 

\end{abstract}

\maketitle

\thispagestyle{fancy}


\section{Introduction}

Narrow Line Seyfert 1 (NLS1) objects are a class of active galactic nuclei
(AGN) discovered by \citet{osterbrock85} and
 identified by their optical properties: narrow permitted lines (FWHM
 (H$\beta$) $<$ 2000 km s$^{-1}$) emitted from the broad line region,
 [OIII]/H$\beta$ $<$ 3, and a bump due to FeII (for a review see
 e.g. \citet{pogge00}). They also exhibit high X-ray variability, steep X-ray spectra and a
prominent soft X-ray excess. These characteristics point to systems with
smaller masses of the central black hole (10$^{6}$-10$^{8}$ M$_\odot$) and
higher accretion rates (up to 90\% of the Eddington value) with respect to
blazars and radio galaxies. Although NLS1s usually do not show strong
radio emission, a small fraction ($\sim$7\%) is radio-loud
\citep{komossa06}. At radio
frequencies, radio-loud NLS1s usually display a relatively compact
morphology without extended structures, and strong and variable
emission with a flat spectrum, suggesting that relativistic jets may
form in these systems. \\ 
The discovery by {\it Fermi}-LAT of
$\gamma$-ray emission in the radio-loud NLS1 PMN\,J0948+0022
\citep{abdo09a} provided a strong support to the presence of a closely
aligned relativistic jet. VLBI observations of PMN\,J0948+0022
pointed out the presence of a bright and compact component with an
inverted spectrum that dominates the radio emission, and a
faint jet feature \citep{giroletti11,foschini11,doi06}. 
The variable emission and
the brightness temperature exceeding the equipartition brightness
temperature indicates that the source must be relativistically
beamed, with Doppler factor $\delta > 1$ \citep{giroletti11}.\\ 
In addition to PMN\,J0948+0022, variable $\gamma$-ray
emission was detected in other 4 NLS1 objects, all radio loud:
1H\,0323+342, PKS\,1502+036, PKS\,2004$-$447 \citep{abdo09b} and more
recently SBS\,0846+513 \citep{donato11}. The increasing number of 
$\gamma$-ray detection of radio-loud NLS1s 
suggests that they form a new class of 
gamma-ray emitting AGNs. This discovery poses 
intriguing questions about our knowledge of the blazar sequence, 
the development of relativistic jets, and the evolution of radio-loud
AGNs. Detailed studies of this
new class of gamma-ray AGNs are fundamental for shedding light on the
characteristics of these objects and the difference with respect to
the other two AGN populations, blazars and radio galaxies, emitting in
the gamma-ray energy band. Redshift and radio luminosity of the
gamma-ray RL-NLS1s are 
reported in Table \ref{prop}.\\

\begin{table}
\begin{center}
\begin{tabular}{lrc}
\hline
Source& z$\;\;\;$ &$\;\;\;$ Log L$_{\rm 1.4 GHz}$ \\
      &   &$\;\;\;$ (W/Hz)\\
\hline
PMN\,J0948+0022& $\;\;\;\;$0.585& $\;\;\;$ 26.19\\
1H\,0323+342   & $\;\;\;\;$0.061& $\;\;\;$ 24.73\\
PKS\,1502+036  & $\;\;\;\;$0.409& $\;\;\;$ 26.35\\
PKS\,2004$-$447  & $\;\;\;\;$0.24 & $\;\;\;$ 26.11\\ 
SBS\,0846+513  & $\;\;\;\;$0.584& $\;\;\;$ 25.71\\
\hline
\end{tabular}
\caption{Redshift and radio luminosity at 1.4 GHz for the gamma-ray NLS1s.}
\label{prop}
\end{center}
\end{table}

\section{Radio data}

Among the 5 gamma-ray emitting NLS1s, we investigated the radio
properties of the three objects lacking detailed studies of their radio
emission: SBS\,0846+513, PKS\,1502+036, and PKS\,2004$-$447. For this
purpose we analysed archival VLA/VLBA data. When
possible, we selected datasets at different epochs and at various
frequencies in order to perform both variability and spectral
studies. In Table \ref{log} we report the information on the datasets
considered in this work.\\
Radio data were retrieved from the archive
and the data reduction was performed with the NRAO \texttt{AIPS}
package. Images were produced in the standard way, 
after a few phase-only self-calibration
iterations (a detailed discussion on the data analysis will be
presented in a forthcoming paper). \\
In the following we present the main results on the radio properties
of SBS\,0846+513, PKS\,1502+036, and PKS\,2004$-$447 obtained from this
analysis. \\

\begin{table}
\begin{center}
\begin{tabular}{crccc}
\hline
Source&Array&Obs. date&Freq&Flux\\
      &     &         &GHz & mJy\\
\hline
SBS\,0846+513& VLA &10 Apr 1986&  5& 286$\pm$11\\
             & VLA &06 Jan 1996&  5& 332$\pm$13\\
             & VLA &09 May 2009&  5& 196$\pm$8\\
             & VLBA&06 Jan 1996&  5& 281$\pm$28\\
             & VLBA&12 Mar 2011&8.4& 210$\pm$21\\
\hline
PKS\,1502+036& VLA &14 Apr 2007& 22& 611$\pm$24\\
             &VLBA &11 Jan 2002& 15& 562$\pm$56\\
             &VLBA &21 Jul 2006& 15& 545$\pm$54\\
\hline
PKS\,2004$-$447&VLBA &13 Oct 1998&1.4& 530$\pm$53\\
\hline
\end{tabular}
\caption{Information on the datasets of the three RL-NLS1s 
discussed in this work, and their flux density.}
\label{log}
\end{center}
\end{table}

\subsection{SBS\,0846+513}

\begin{figure}[t]
\includegraphics[width=80mm]{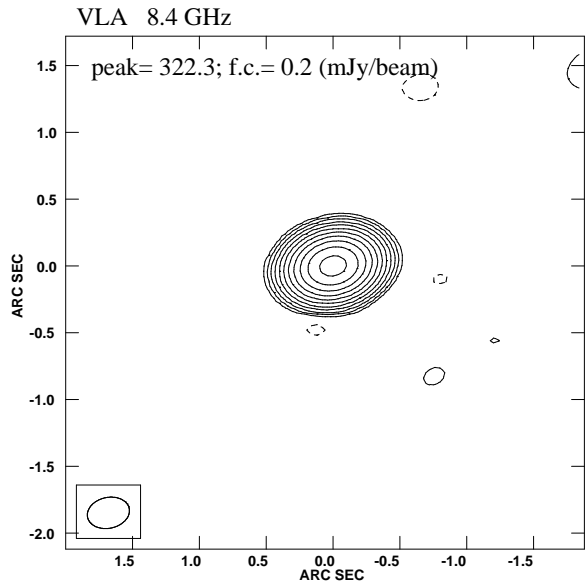}
\caption{VLA image at 8.4 GHz of SBS\,0846+513. On the image we 
provide the restoring beam, plotted in the bottom left corner, 
the peak flux density in mJy/beam, and the first contour (f.c.)
intensity in mJy/beam, which is 3 times the off-source noise level. 
Contour levels increase by a factor of 2.} 
\label{FMJ-f1}
\end{figure}

\begin{figure}[t]
\centering
\includegraphics[width=80mm]{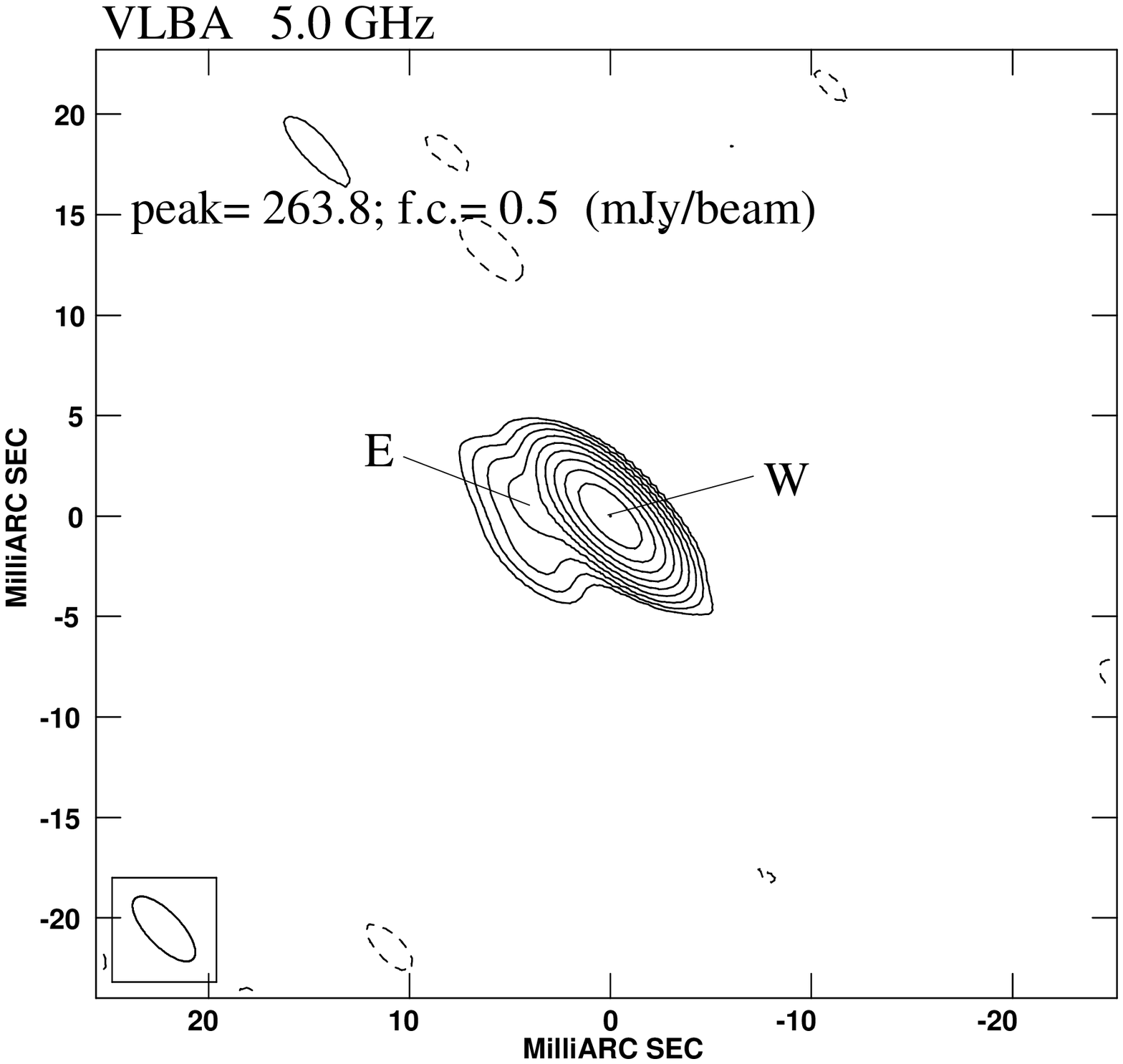}
\caption{VLBA image at 5.0 GHz of SBS\,0846+513. On the image we 
provide the restoring beam, plotted in the bottom left corner, 
the peak flux density in mJy/beam, and the first contour (f.c.)
intensity in mJy/beam, which is 3 times the off-source noise level. 
Contour levels increase by a factor of 2.} 
\label{FMJ-f2}
\end{figure}

\begin{figure}[t]
\centering
\includegraphics[width=80mm]{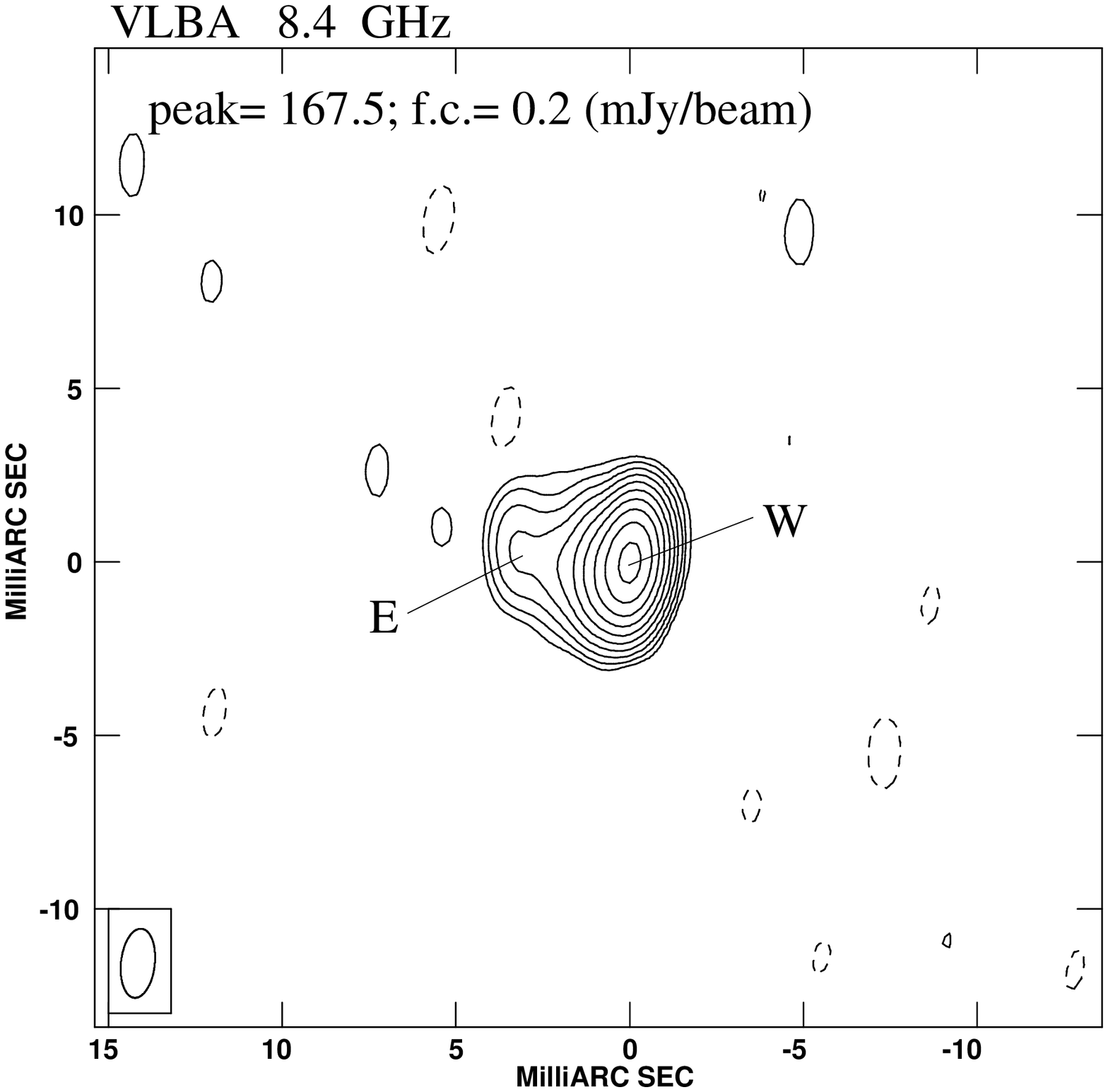}
\caption{VLBA image at 8.4 GHz of SBS\,0846+513. On the image we 
provide the restoring beam, plotted in the bottom left corner, 
the peak flux density in mJy/beam, and the first contour (f.c.)
intensity in mJy/beam, which is 3 times the off-source noise level. 
Contour levels increase by a factor of 2.} 
\label{FMJ-f3}
\end{figure}

SBS\,0846+513 is the most recent NLS1 detected in gamma rays by 
{\it Fermi}-LAT \citep{donato11}. 
Multi-epoch studies of the radio emission from the source indicate 
that the flux density has some degrees of variability 
(i.e. a variation of 70\% at 4.8 GHz during 1986-2009).
The resolution provided by the VLA is not adequate to resolve the source
structure (Fig. \ref{FMJ-f1}). When imaged with the high spatial resolution
provided by the VLBA, the source shows two
asymmetric components resembling a core-jet structure
(Figs. \ref{FMJ-f2} and \ref{FMJ-f3}),  
as also pointed
out in a previous work by \citet{taylor05}. The flux density ratio
between components E and W is about 19 and 14 at 5 GHz and 8.4 GHz,
respectively. They are separated by about 3.5 mas 
($\sim$23 pc at source redshift z=0.5835). 
Although the source is unresolved with
the VLA, pc-scale VLBA observations cannot recover all the VLA flux
density. Such a discrepancy between the flux
densities may be related to the intrinsic variability of the source. 
Even when VLA and
VLBA observations were performed simultaneously (i.e. 1996, January 6), the
VLBA flux density was only 85\% of the value measured with the VLA at
the same frequency. 
This suggests that the missing flux on parsec scale may be related to 
extended, low-brightness features, 
like a jet structure, that is resolved out by the VLBA array. For more
details see \citet{dammando12}.

\subsection{PKS\,1502+036}

\begin{figure}[t]
\centering
\includegraphics[width=80mm]{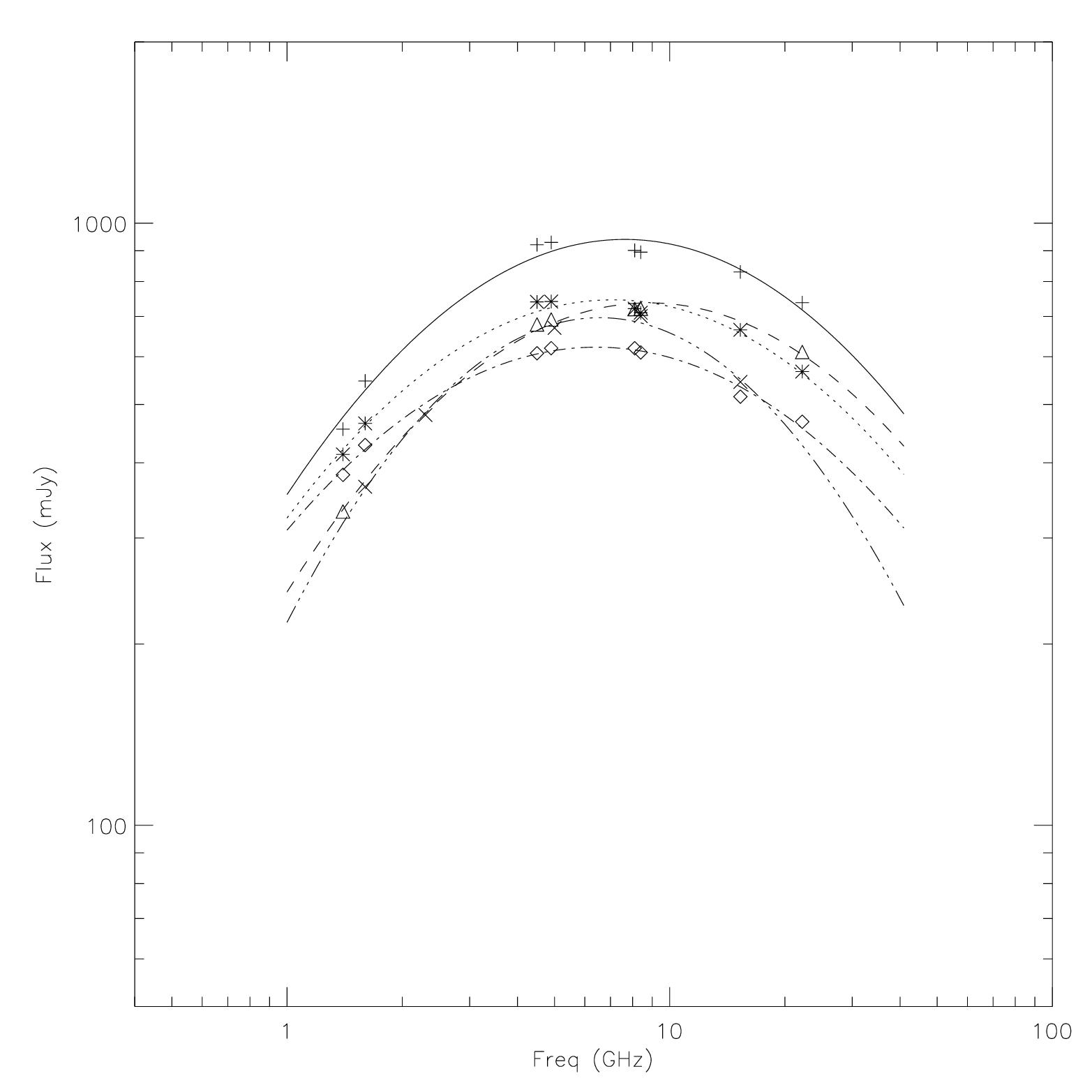}
\caption{Multi-epoch radio spectra of PKS\,1502+036 observed with VLA and VLBA 
during 5 observing runs: + and solid line = 25 September 1999 (VLA); 
asterisk and dot line = 03 July 2002 (VLA); diamonds and dot-dash 
line = 13 September 2003 (VLA); X and dash-3dots line = 21 July 2006 (VLBA); 
triangles and dashed line = 14 April 2007 (VLA). VLA data are from
\citet{dd00,tinti05,mo07}.}  
\label{FMJ-f4}
\end{figure}

\begin{figure}[b]
\centering
\includegraphics[width=80mm]{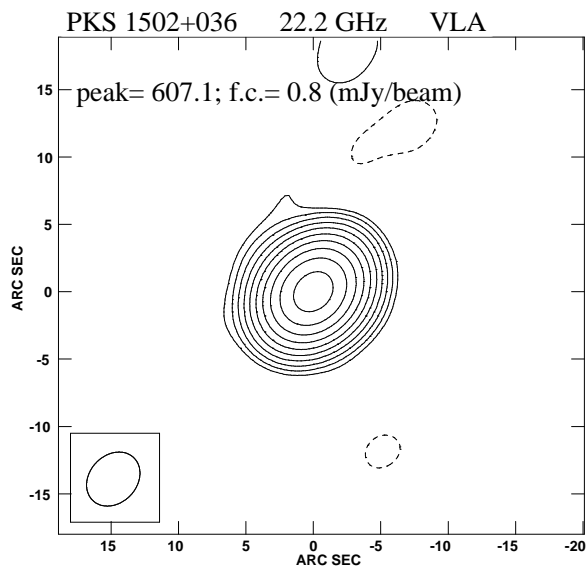}
\caption{VLA image at 22.2 GHz of PKS\,1502+036. On the image we 
provide the restoring beam, plotted in the bottom left corner, 
the peak flux density in mJy/beam, and the first contour (f.c.)
intensity in mJy/beam, which is 3 times the off-source noise level. 
Contour levels increase by a factor of 2.} 
\label{FMJ-f5}
\end{figure}

\begin{figure}[t]
\centering
\includegraphics[width=80mm]{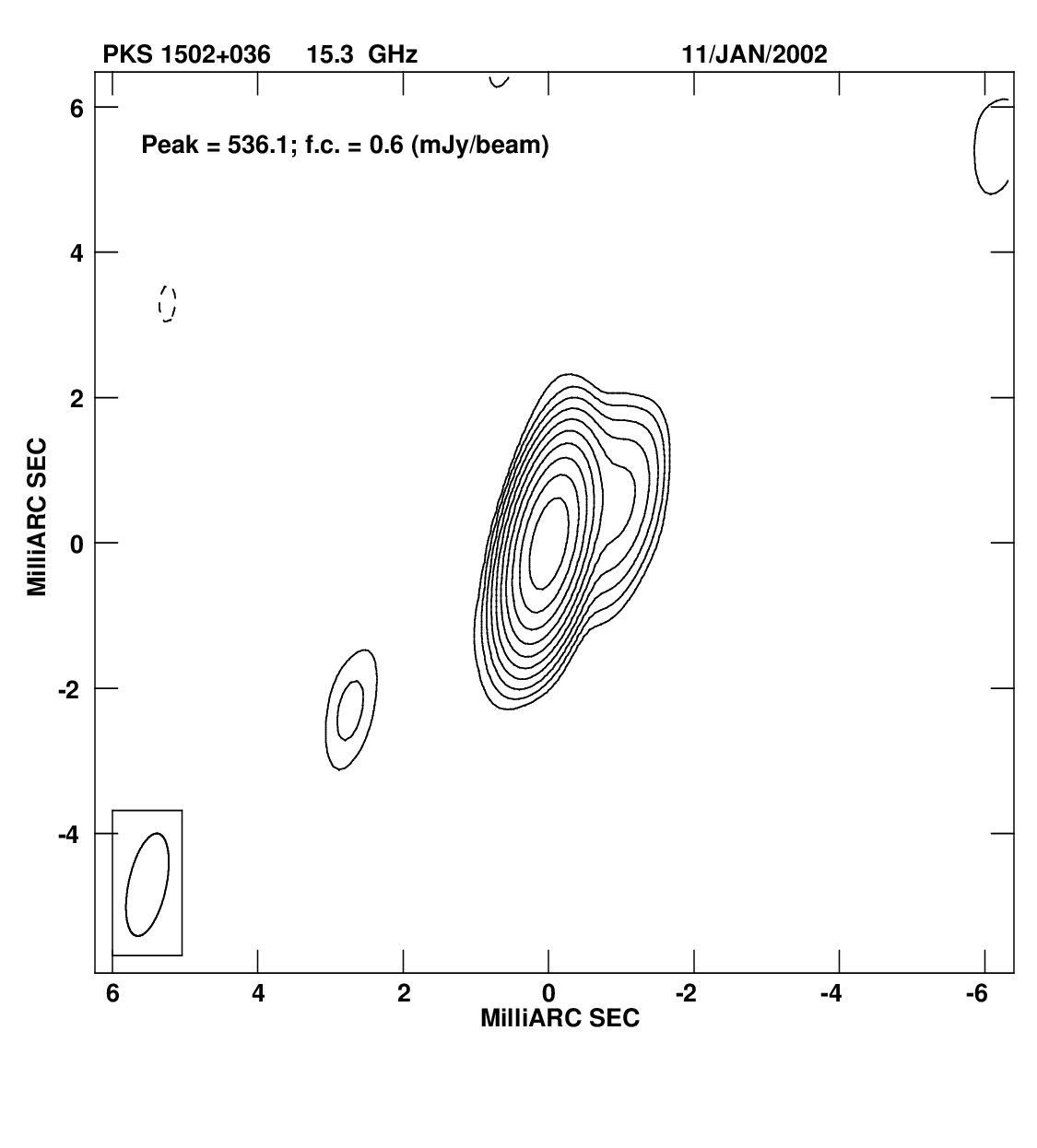}
\caption{VLBA  image at 15.3 GHz of PKS\,1502+036 collected on 11
  January 2002. On the image we 
provide the restoring beam, plotted in the bottom left corner, 
the peak flux density in mJy/beam, and the first contour (f.c.)
intensity in mJy/beam, which is 3 times the off-source noise level. 
Contour levels increase by a factor of 2.} 
\label{FMJ-f6}
\end{figure}

\begin{figure}[b]
\centering
\includegraphics[width=80mm]{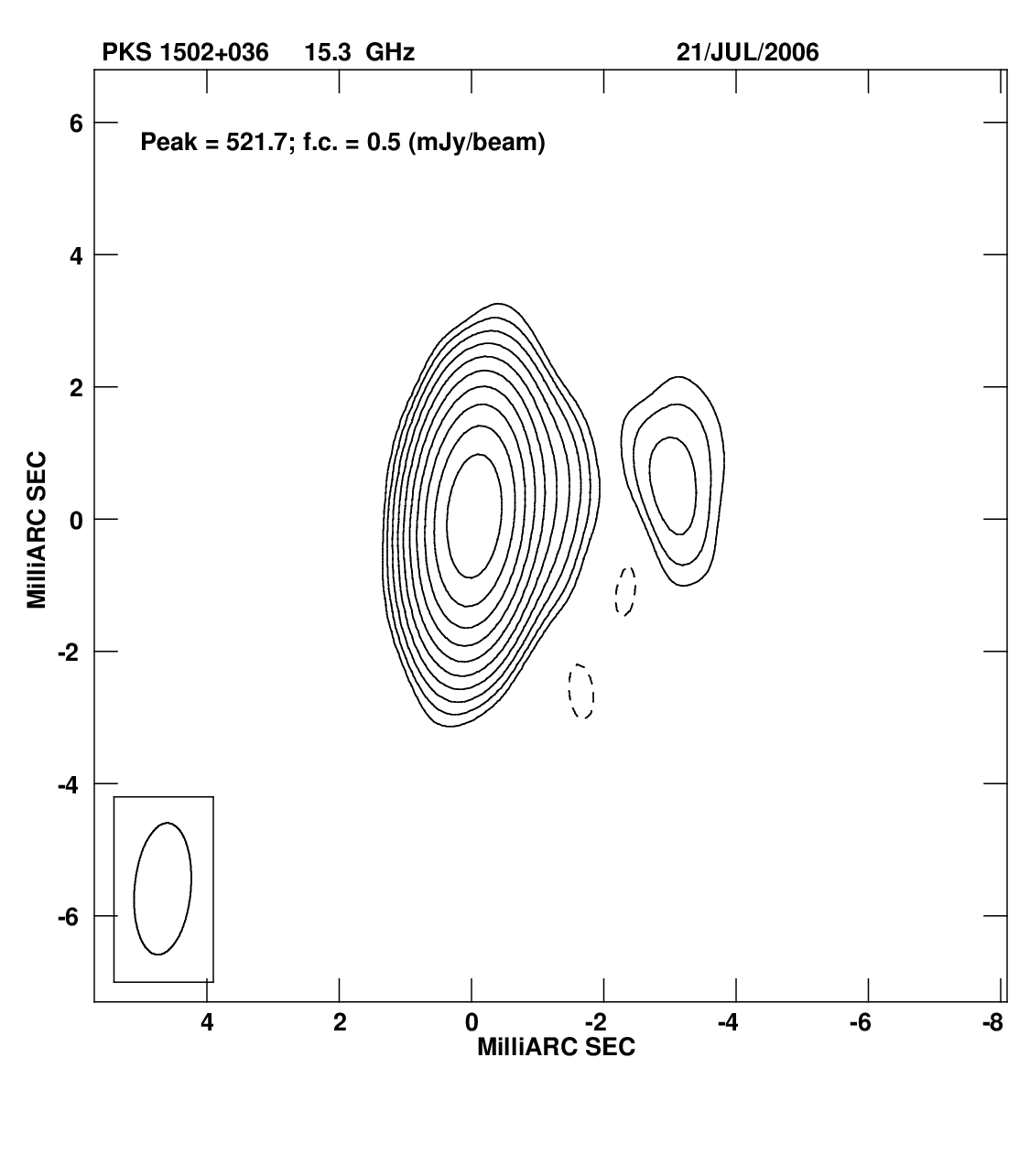}
\caption{VLBA  image at 15.3 GHz of PKS\,1502+036 collected on 21 July
  2006. On the image we 
provide the restoring beam, plotted in the bottom left corner, 
the peak flux density in mJy/beam, and the first contour (f.c.)
intensity in mJy/beam, which is 3 times the off-source noise level. 
Contour levels increase by a factor of 2. } 
\label{FMJ-f7}
\end{figure}

With its convex radio spectrum peaking above 5 GHz,
PKS\,1502+036 was selected by \citet{dd00} as part 
of the bright sample of high frequency peakers (HFPs). 
Given the anticorrelation found
between the intrinsic source size (i.e. age) and the spectral peak
\citep{odea97}, HFP objects should represent radio sources in
the very earliest stages of their radio evolution. However, by the
analysis of their radio properties (i.e. variability, morphology,
polarization), it turned out that a great fraction ($\sim$60\%) 
of objects from the
bright HFP sample are contaminant blazars rather than genuinely young
radio sources \citep{mo08}. 
Based on simultaneous multifrequency 
(from 1.4 to 22 GHz) VLA observations carried out in several epochs, 
PKS\,1502+036 turned out to
possess strong spectral and flux density variability
(Fig. \ref{FMJ-f4}). 
During these epochs the spectral peak was at 7.6, 7.1, 6.4, 6.5, and 8.8
GHz, respectively.
Variability is a typical property displayed by blazars, 
while young radio sources are non-variable objects. For this reason
PKS\,1502+036 was rejected from the HFP sample of young radio sources
\citep{mo08}. \\
PKS\,1502+036 is unresolved on the typical VLA
scales (Fig. \ref{FMJ-f5}). 
However, when imaged with the parsec scale resolution 
provided by VLBA observations, its radio structure is marginally
resolved and a second component seems to emerge from the core. 
In the VLBA images at 15 GHz the double radio structure is clearly 
resolved suggesting a core-jet structure. The radio emission is
dominated by the core component, while the jet-like feature is only
4\% of the total flux density.
Comparing the 15-GHz observations carried out at two different 
epochs (11 January 2002, and 21 July 2006, Figs. \ref{FMJ-f6} and
\ref{FMJ-f7}) it seems
that in the second-epoch image there is a weak component of 4 mJy, 
separated by 3.05 mas ($\sim$16.5 pc at z=0.4088) from the core which was
not detected in the previous observations, although the sensitivity
between the two observations is similar.

\subsection{PKS\,2004$-$447}

\begin{figure}[t]
\centering
\includegraphics[width=80mm]{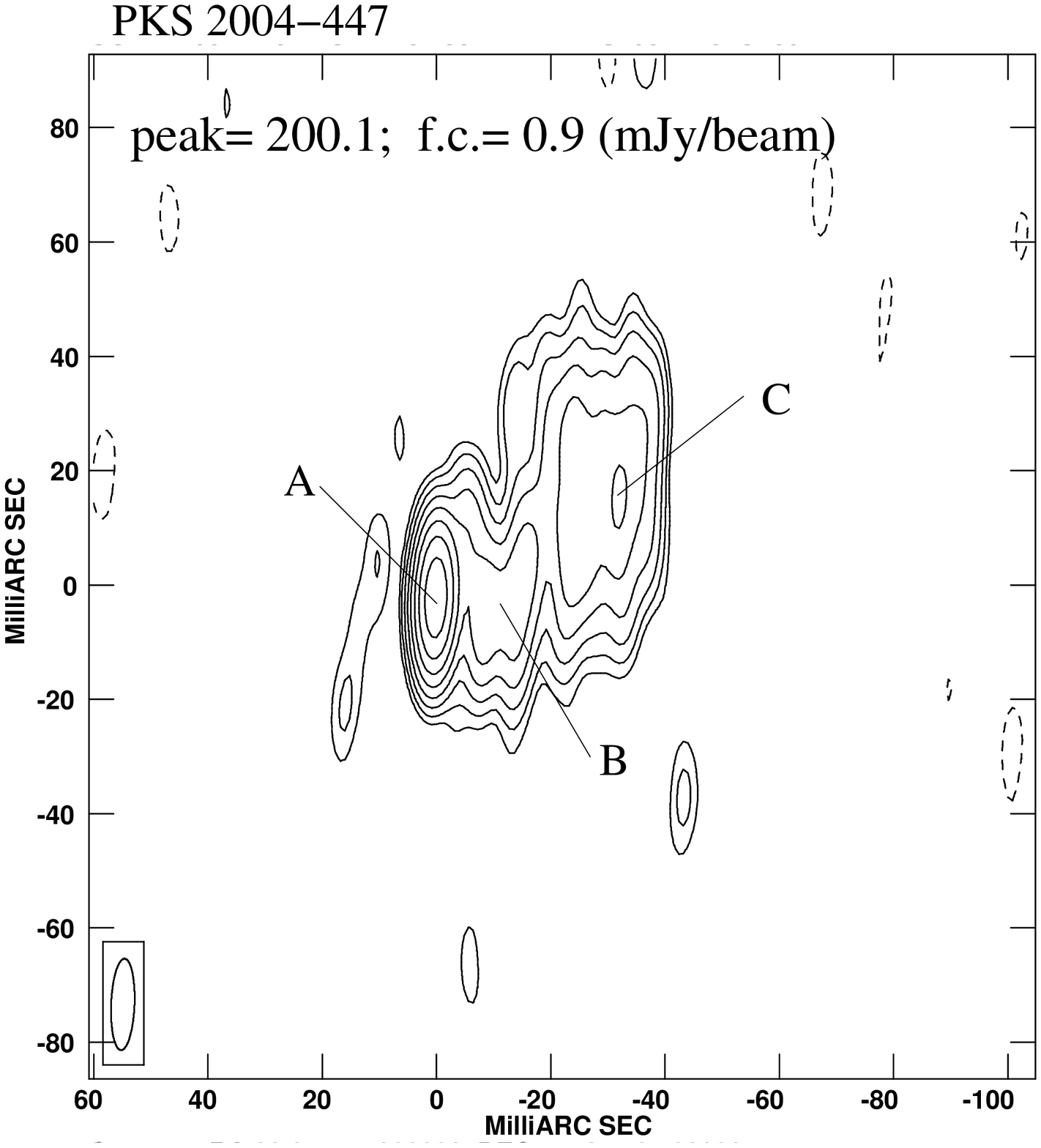}
\caption{VLBA  image at 1.4 GHz of PKS\,2004-447. On the image we 
provide the restoring beam, plotted in the bottom left corner, 
the peak flux density in mJy/beam, and the first contour (f.c.)
intensity in mJy/beam, which is 3 times the off-source noise level. 
Contour levels increase by a factor of 2.} 
\label{FMJ-f8}
\end{figure}

PKS\,2004$-$447 is a powerful ($L_{\rm 1.4 GHz} \sim 1.3 \times 10^{26}$ W/Hz) 
radio source at redshift z=0.24 and it is one of the four radio-loud
NLS1 detected in gamma rays during the first year of {\it Fermi}
operation \citep{abdo09b}. 
However its classification is still uncertain. \citet{oshlack01} 
classified PKS\,2004$-$447 as a genuine radio-loud NLS1 based on the
optical definition. At radio wavelengths it is characterized by a steep
($\alpha$ $>$ 0.5, S($\alpha$) $\propto$ $\nu^{- \alpha}$) synchrotron 
spectrum at high frequency (above 8.4 GHz) and the absence of 
significant flux density variability. When imaged with
arcsecond-resolution (e.g. ATCA) it is unresolved. Based on these 
characteristics \citet{gallo06} proposed this source as a genuine 
compact symmetric object (CSO). On the other hand, PKS\,2004$-$447 has 
been included in the CRATES catalog of flat spectrum objects 
\citep{healey07} due to its flatter spectrum ($\alpha \sim$0.3) 
below 4.8 GHz, questioning the nature of the radio emission. 
Indeed a flat spectrum is usually an indication of a self-absorbed 
component as in the blazar population, where the emission is dominated 
by the core region enhanced by projection effects. However, 
a flat spectrum may also be an indication of a convex spectrum 
as those found in young radio objects, 
whose spectral peak occurs between 
the frequencies considered.\\
So far the only information on the pc-scale structure of PKS\,2004$-$447 
is from archival VLBA data at 1.4 GHz (Fig. \ref{FMJ-f8}). The radio structure 
has an angular size of 45 mas (170 pc) with a position angle of 
$-$50$^{\circ}$, and it is resolved into 3 main components. 
The lack of the spectral index information does not allow us 
to unambiguously classify the sub-components. For example the compact 
component A can be either the core of a core-jet boosted blazar, 
or a bright hotspot of an asymmetric young radio source.

\section{Discussion and future work}

Relativistic jets are the most extreme manifestation of the power
generated by a super massive black hole (SMBH) in the center of an
AGN, with a large fraction of this power being emitted in the
gamma-ray energy band. 
Gamma-ray emission from radio-loud NLS1s provides evidence that
relativistic jets may be not a peculiarity of blazars and
radio galaxies/quasars, both hosted in giant elliptical galaxies. 
NLS1s are generally hosted in spiral galaxies \citep{deo06,zhou06},
and the presence of relativistic jets in these objects challenges the
paradigm that the relativistic jets can form only in elliptical
\citep{marscher09}. However, only for one out of the five gamma-ray
NLS1s, 1H\,0323+342, a high resolution HST observation suggests that
this AGN is hosted in a spiral galaxy \citep{zhou07,anton08}. 
Multi-band campaigns
aimed at studying the Spectral Energy Distribution (SED) 
of this possible new gamma-ray emitting
population pointed out that their physical properties and jet power
are similar to those found in blazars. In
particular the jet power estimated for PMN\,J0948+0022 and
PKS\,1502+036 is in the region of flat spectrum radio quasars
(FSRQ), while in the case of 1H\,0323+342 and PKS\,2004$-$447 
the jet power is in the range typical of BL Lac objects
\citep[see][]{abdo09b}. \\
The blazar-like behaviour of these objects
casts doubts questioning their nature and role in the context of the
blazar sequence.
Considering the relative small black hole masses
and the high accretion rates, RL-NLS1s allow us to extend the
physical properties of relativistic objects to low-mass systems.
However, models developed so far can fit the SED only from the IR to
shorter wavelengths, leaving the low-energy part of their spectrum not
fully understood.\\

From the preliminary analysis of the radio data presented here, 
we found that all the sources are compact on kpc scales, but they are 
resolved on parsec scales. 
The radio emission of SBS\,0846+513 and PKS\,1502+036
is dominated by a flat-spectrum compact and bright component from
which a
fainter jet-like
feature emerges. Both sources show substantial flux density variability.\\
In the case of PKS\,1502+036, the availability of
two-epoch VLBA observations at 15 GHz allowed us to detect a new
component at about 3 mas from the central region that was not visible
in the first observing epoch. Assuming that this component is the same
knot of the jet detected at about 1 mas during the first epoch, we
estimate an apparent superluminal expansion velocity of $\sim$8c,
providing further evidence for boosting effects. However, with only two
observations taken with a time separation of 4 years,
we cannot unambiguously state that the two components are
the same one at the different epochs. To confirm this result we are now
analyzing additional archival VLBA observations which allow a better
time sampling. \\ 
The interpretation of the radio properties of PKS\,2004$-$447 is more
uncertain due to the lack of spectral information. In fact, the bright
and compact component at the easternmost edge of the source 
may be either the source
core, or a very compact hot-spot like those found in a few young radio
sources \citep[e.g.][]{mo06}. The little flux density variability reported by
\citet{gallo06}, and its steep spectrum above 8.4 GHz make
PKS\,2004-447 a candidate of being a young radio source rather than a
blazar. To unveil the nature of this source new multifrequency
observations with parsec-scale resolution have been requested.\\

Although the radio properties are important to constrain the
emission mechanisms in the low-energy part of the spectrum of
RL-NLS1s, at
centimeter wavelengths, i.e. those discussed here, their emission is
highly affected by synchrotron self-absorption. The one-zone
homogeneous model adopted for the broad-band SED of these sources
\citep[e.g.][]{abdo09b} is based on a single
synchrotron emission component which does not fit the radio data,
being all the emission below 10$^{12}$ Hz considered self-absorbed. Information
on the mm/sub-mm emission is fundamental to set tight constraints on
the low-energy emission, and ALMA observations will open a new window
for investigating these sources in an unexplored frequency domain.\\

\begin{acknowledgments}

The {\it Fermi} LAT Collaboration acknowledges support from a number
of agencies and institutes for both development and the operation of
the LAT as well as scientific data analysis. These include NASA and
DOE in the United States, CEA/Irfu and IN2P3/CNRS in France, ASI and
INFN in Italy, MEXT, KEK, and JAXA in Japan, and the K.~A.~Wallenberg
Foundation, the Swedish Research Council and the National Space Board
in Sweden. Additional support from INAF in Italy and CNES in France
for science analysis during the operations phase is also gratefully
acknowledged. The VLA and the VLBA are operated by the US National
Radio Astronomy Observatory which is a facility of the National
Science Foundation operated under cooperative agreement by Associated
Universities, Inc. This research has made use of the NASA/IPAC
Extragalactic Data base (NED) which is operated by the Jet Propulsion
Laboratory, California Institute of Technology, under contract with
the National Aeronautics and Space Administration. 

\end{acknowledgments}

\bigskip 

\end{document}